\documentclass{article}
\usepackage{spconf,graphicx}
\usepackage{amsmath,amssymb,amsfonts}
\usepackage{multirow}
\usepackage{booktabs}
\usepackage{xcolor,colortbl}
\usepackage{dblfloatfix}
\usepackage{caption}
\usepackage[noadjust]{cite}
\usepackage{hyperref}

\usepackage{multirow}
\usepackage{tabularx}

\usepackage{dblfloatfix}

\title{CryCeleb: A Speaker Verification Dataset Based on Infant Cry Sounds}

\name{David Budaghyan$^1$, Charles C. Onu$^{1, 2, 5}$, Arsenii Gorin$^1$, Cem Subakan$^{2, 3, 4}$, Doina Precup$^{2, 5, 6}$}

\address{$^1$Ubenwa Health, $^2$Mila-Quebec AI Institute, $^3$Université Laval, \\ $^4$Concordia University, $^5$McGill University, $^6$DeepMind}

\newcommand{\cem}[1]{#1}

\newlength{\bibitemsep}
\newlength{\bibparskip}
\let\oldthebibliography\thebibliography
\renewcommand\thebibliography[1]{%
  \oldthebibliography{#1}%
  \setlength{\parskip}{\bibitemsep}%
  \setlength{\itemsep}{\bibparskip}%
}

\begin{document}

\maketitle


\begin{abstract}

This paper describes the Ubenwa CryCeleb dataset - a labeled collection of infant cries - and the accompanying CryCeleb 2023 task, which is a public speaker verification challenge based on cry sounds. 
We released more than 6 hours of manually segmented cry sounds from 786 newborns for academic use, aiming to encourage research in infant cry analysis. 
The inaugural public competition attracted 59 participants, 11 of whom improved the baseline performance. 
The top-performing system achieved a significant improvement scoring 25.8\% equal error rate, which is still far from the performance of state-of-the-art adult speaker verification systems. Therefore, we believe there is room for further research on this dataset, \cem{and also} potentially extending beyond the verification task.

\end{abstract}


\section{Introduction}

Clinical research on the analysis of infant cries goes back to the 1960s~\cite{wasz1985twenty}. 
These days, machine learning techniques are demonstrating promising results in cry-based detection of cry reasons (hunger, pain, etc) and, more importantly, health pathologies, such as neurological injury~\cite{ji2021review,parga2020defining,gorin2023self}.

When deployed in hospitals or households with multiple babies, a cry analysis system should be able to identify the infant associated with the cry. Training such a model requires data with multiple recordings per infant. Given the complexity of data collection from newborns, such resources are extremely scarce. The most popular and diverse Chillanto database has 127 newborns but only one recording per infant~\cite{reyes2004system}.

In this work, we present the Ubenwa CryCeleb dataset, a first-of-its-kind collection of cries labeled with anonymized infant identities. Comprising 786 infants and 6.5 hours of cry expirations, with 348 infants recorded at least twice in different time frames (right after birth and pre-discharge from hospital), we aim to foster research in cry verification and, more broadly and importantly, to advance the field of infant cry analysis using verification as a proxy task. The dataset is available online\footnote{\url{huggingface.co/datasets/Ubenwa/CryCeleb2023}} under Creative Commons license.
In addition, we report on the first public baby verification competition and summarize the results.

\section{Dataset description}

CryCeleb2023 is a curated and anonymized subset of a large clinical study conducted by Ubenwa Health. In the next section, we describe the data collection and pre-processing steps. 

\subsection{Data Preparation}
The original cry recordings were made either within an hour of birth or upon discharge from the hospital (typically within 24 hours of birth up to a few days). 
The cries were collected by medical personnel using the Ubenwa study application~\cite{onu2017ubenwa} and a Samsung A10 smartphone held at 10-15 cm from the newborn's mouth. The original audio files, recorded at 44.1 kHz, were  downsampled to 16 kHz and stored as wav PCM.

Each recording was manually segmented by a human annotator into `expiration', `inspiration' or `no cry' segments. 
The CryCeleb dataset consists solely of the expiration segments, which we refer to as cry sounds. Inspirations (breath) are excluded as they are generally too short, hard to detect, and less likely to convey information about the vocal tract. Also, we manually removed any cry sounds containing personally identifiable information, such as human speech.

\subsection{Metadata and Descriptive Statistics}

This section summarizes the information about audio files included in the dataset and the associated anonymized metadata available for download.
Table~\ref{tab:dataset_stats} provides general statistics of the dataset.

\begin{table}[h]
\centering
\begin{tabular}{l|r}
\hline
Number of cry sounds (expirations)         & 26093 \\
Number of original recordings & 1372  \\
Number of infants             & 786   \\
Total cry time (minutes)      & 391  \\
\hline
\end{tabular}
\caption{Summary statistics of the dataset.}
\label{tab:dataset_stats}
\end{table}

The audio is accompanied by a metadata file with fields summarized in Table~\ref{tab:metadata_fields}. The 26093 rows of csv file provide complete information about the cry audio files.

\begin{table}[h]
\centering
\begin{tabular}{r|l}
\hline
\multicolumn{1}{c|}{\textbf{Field}} & \multicolumn{1}{c}{\textbf{Description}} \\
\hline
baby\_id            & Unique infant identifier. \\
period              & Recording time (birth or discharge). \\
duration     & Length of cry sound in seconds. \\
split               & Split for the challenge. \\
chrono\_idx & Chronological ordering of cries. \\
file\_name          & Path to cry sound. \\
file\_id            & Cry sound unique identifier. \\
\hline
\end{tabular}
\caption{Metadata fields.}
\label{tab:metadata_fields}
\end{table}

\begin{figure}[t]
\centering
\includegraphics[width=1.0\columnwidth]{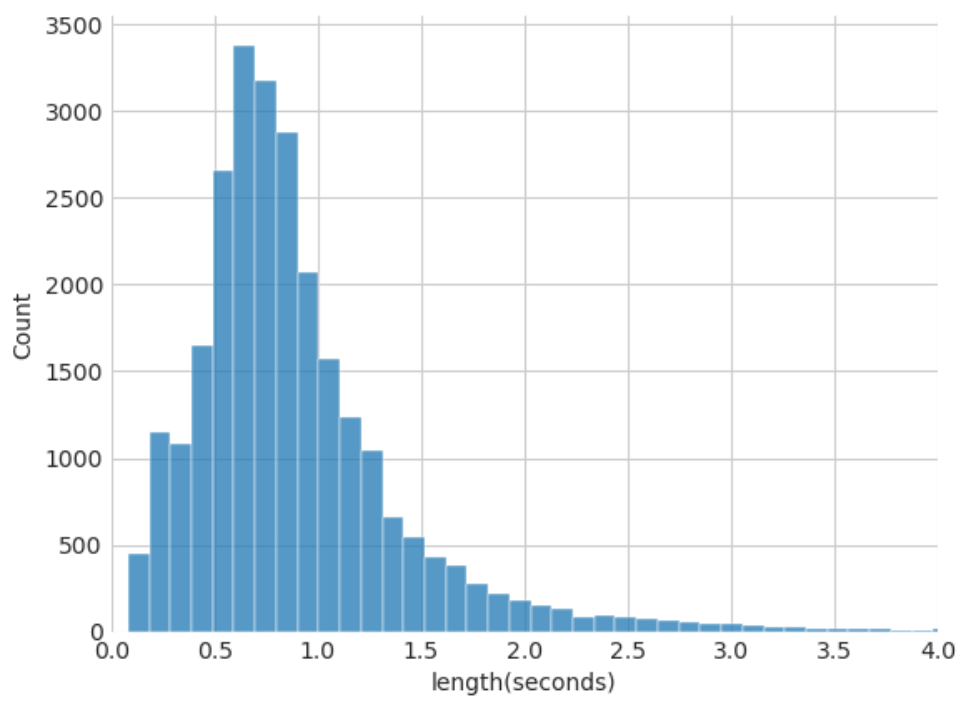}
\begin{minipage}[t]{0.9\columnwidth}
\caption{Histogram of cry sound durations.}
\label{fig:hist-len}
\end{minipage}
\end{figure}

\begin{figure}[t]
\centering
\includegraphics[width=1.0\columnwidth]{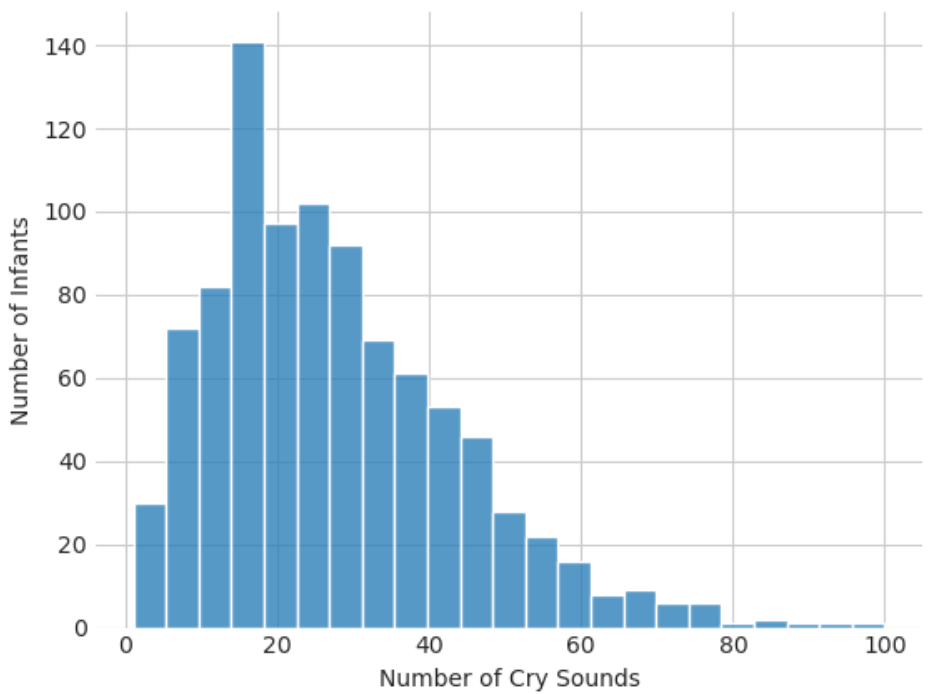}
\begin{minipage}[t]{0.9\columnwidth}
\caption{Number of infants per number of cry sounds.}
\label{fig:hist-numinf}
\end{minipage}
\end{figure}

Figures~\ref{fig:hist-len} and~\ref{fig:hist-numinf} show the distribution of cry sound duration and number of cries per infant. Most of the cry sounds are short (0.5 - 1.0 seconds) with only 0.3\% of expirations longer than 4 seconds. At the same time, there are multiple cry sounds corresponding to each infant. However, cry sounds (expirations) collected within one recording session, tend to have similar acoustic characteristics.

\section{CryCeleb 2023 Challenge}

\begin{figure}[h]
\centering
\includegraphics[width=0.9\columnwidth]{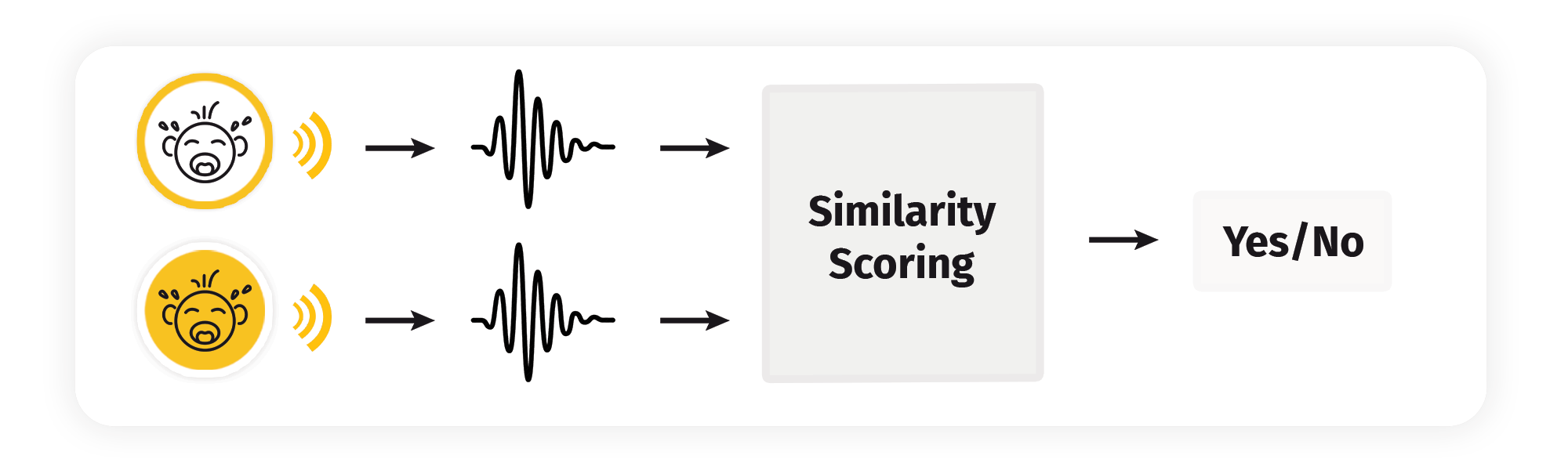}
\begin{minipage}[t]{0.9\columnwidth}
\caption{CryCeleb challenge verification task. Given two recordings, predict if they belong to the same infant}
\label{fig:verification}
\end{minipage}
\end{figure}

CryCeleb 2023 was a two-month \cem{long} machine learning competition\footnote{\url{huggingface.co/spaces/competitions/CryCeleb2023}} where contestants were asked to develop a system capable of determining whether two distinct cry recordings originated from the same infant (see Figure~\ref{fig:verification}). 

\subsection{Challenge design details}

The system should analyze any pair of cries and assign a similarity score to determine if the two sounds belong to the same baby, with an ideal system always assigning higher scores to positive pairs (two cry sounds from the same infant) than to negative pairs. 
For decision-making, a threshold can be applied to the assigned scores. If a score is greater than the threshold, it will indicate that the system accepts the two cries as belonging to the same infant. 

Submissions are ranked using the Equal Error Rate (EER)~\cite{brummer2006application}. The EER is the point on the ROC curve at which the false acceptance rate equals the false rejection rate. 
The lower the EER, the better.

For the CryCeleb2023 challenge, we have partitioned all infants into three sets: train (586 infants), dev (40 infants), and test (160 infants). All infants in the dev and test sets have recordings from both the birth (B) and discharge (D) periods. This is not true for all infants in the train set.

\begin{table}[h]
\begin{center}
\begin{tabular}{|c|c|c|c|}
\hline
 & \multicolumn{3}{c|}{\textbf{Split}} \\
\cline{2-4}
\textbf{Time(s) of Recording(s)} & \textbf{\textit{train}}& \textbf{\textit{dev}}& \textbf{\textit{test}} \\
\hline
Both birth and discharge & 348 & 40 & 160  \\
\hline
Only birth & 183 & 0 & 0  \\
\hline
Only discharge & 55	 & 0 & 0  \\
\hline
\multicolumn{1}{c|}{} & 586 & 40 & 160 \\
\cline{2-4}
\multicolumn{4}{l}{}
\end{tabular}
\vspace{-10pt}
\caption{Number of infants by split and recording period.}
\label{tab:split}
\end{center}
\end{table}

To tune the algorithms, participants were provided with suggested development pairs - cross-product of the birth and discharge recordings (all possible combinations) of the dev infants. Similarly, test pairs consist of the $B\times D$ cross product for infants in the test set. Test pairs were provided without answers during the competition. After the public challenge ended, all answers were released to encourage further research, evaluating the full set of 200 identities.

\begin{table}[h]
\centering
\begin{tabular}{l|c|c|c}
\hline
\textbf{Split} & \textbf{\# of +ive pairs} & \textbf{\# of -ive pairs} & \textbf{Total \# of pairs} \\
\hline
dev            & 40                            & 1540                           & 1580 \\
test           & 160                           & 25440                          & 25600 \\
\hline
\end{tabular}
\caption{Number of pairs in dev and test.}
\label{tab:positive_negative_total_pairs}
\end{table}

Each verification pair in both dev and test sets comprises one birth and one discharge recording. 
Pairing different recordings rather than cry sounds from the same recording is more representative of real-world applications for such a verification system, which may involve verifying an infant over multiple days. Additionally, we observed that verifying separate segments from the same recording is easier, possibly because an infant exhibits consistent traits within a single crying "episode" but not across different episodes. 

It's important to emphasize that the dev and test infants were not chosen randomly. Instead, they were randomly sampled from the top 200 infants with the highest cosine similarities between their birth and discharge embeddings, as calculated using the initial non-fine-tuned baseline model described in Section 3.2 and the first row of Table~\ref{tab:eer_baselines}. We opted for these relatively easier pairs due to the difficulty in recognizing an infant in an unseen recording within this dataset. By selecting easier-to-verify pairs, our aim was to add variance to the leaderboard and make the challenge more engaging.

For the competition, the test set was further subdivided into public (1024 pairs) and private (24576 pairs) parts at random. The participants could score their system on a public subset throughout the competition with a limit of 3 times per day. For the final evaluation, participants were asked to provide the top 5 performing systems for evaluation on the private subset.

\subsection{Baselines}
The baseline is a state-of-the-art ECAPA-TDNN speaker verification model~\cite{desplanques2020ecapa} that uses cosine similarity scoring of audio embeddings produced by a neural network encoder.
First, the ``naive'' baseline is pre-trained using a large adult speaker verification corpus - VoxCeleb~\cite{Chung18b} without any adaptation on cry data. We refer to the open-source SpeechBrain implementation~\cite{speechbrain} for further details with the model available in Hugging Face~\cite{voxcelebModel}. As it does not use any fine-tuning on cry data, the performance is quite low - 37.9\% and 38.1\% EER on the dev and test pairs respectively.

Second, the VoxCeleb model \cem{(the baseline described above)} is fine-tuned on CryCeleb training data, specifically focusing on the 348 infants with both birth and discharge recordings (Table~\ref{tab:split}, top left). By limiting the dataset to this subset, we can train the model on all birth recordings while reserving the discharge recordings for validation. This approach enables us to assess the model's ability to generalize patterns learned from birth recordings to discharge recordings, in some sense simulating the verification setting. 
Alternatively, we could have fine-tuned the model on both birth and discharge recordings from the 348 infants or even expanded it to all recordings from the 586 train infants. The former option introduces more data but removes the ability to validate the classification performance. The latter allows for even more data, however, it also increases the number of classes, which could hinder the model's learning. 

The model is trained on 3-5 second random chunks from concatenated cry sounds at each iteration. The best 5 epochs, determined by validation accuracy, are saved, and these 5 checkpoints are then evaluated on the dev pairs using the EER (verification task). The checkpoint with the lowest EER is chosen as our final fine-tuned model. The fine-tuned model achieves an EER of 27.8\% on the dev set and 28.2\% on the test set. It is open-sourced\footnote{\url{huggingface.co/Ubenwa/ecapa-voxceleb-ft2-cryceleb}} along with fine-tuning code\footnote{\url{github.com/Ubenwa/cryceleb2023}}.

\subsection{Baseline performance}

Table~\ref{tab:eer_baselines} summarizes the performance of two baselines on dev and test sets and the following section provides more details about these systems.

\begin{table}[ht]
\centering
\begin{tabular}{l|c|c}
\hline
\textbf{ECAPA-TDNN baseline} & Dev & Test\\
\hline
VoxCeleb pre-training (no fine-tuning) & 37.9\% & 38.1\%\\
 + CryCeleb fine-tuning & 27.8\% & 28.2\% \\
\hline
\end{tabular}
\caption{Performance (EER) of baseline models. First row indicates the performance obtained with the model pre-tained on VoxCeleb and the second row indicates the performance where the first model is fine-tuned on CryCeleb.}
\label{tab:eer_baselines}
\end{table}

\begin{figure*}[ht]
\centering
\begin{minipage}[b]{0.45\textwidth}
  \centering
\centerline{\includegraphics[width=\linewidth]{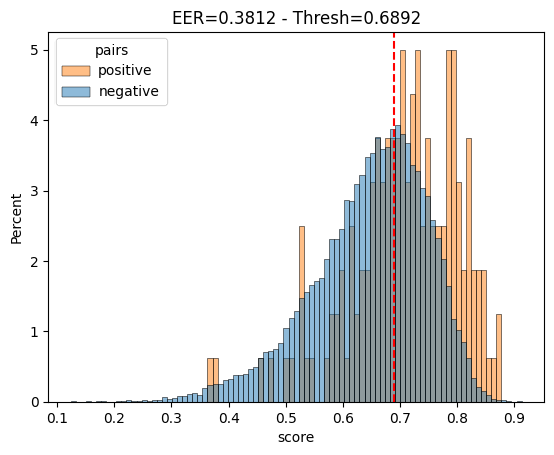}}
\end{minipage}
\begin{minipage}[b]{0.49\textwidth}
\centering

\centerline{
\vspace{-6pt}
\includegraphics[width=0.92\linewidth]{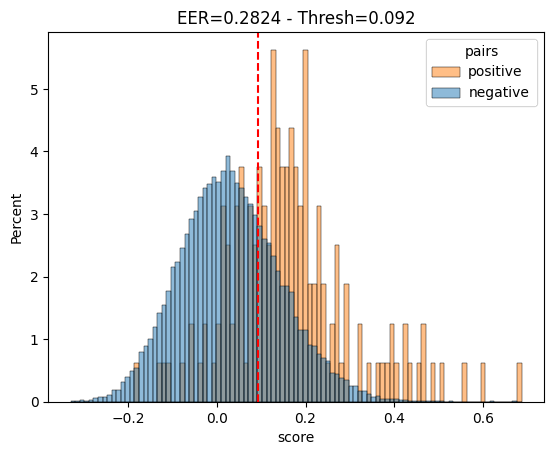}}
\medskip
\end{minipage}
\vspace{-10pt}
\caption{Verification scores for negative and positive pairs produced by the ECAPA-TDNN pre-trained on speech data VoxCeleb (left) and fine-tuned on CryCeleb (right). Classification threshold (red dashed line) is selected to minimize EER.}
\label{fig:hist-perf}
\end{figure*}

\begin{figure*}[h]
\centering
\vspace{-10pt}
\includegraphics[width=2.0\columnwidth]{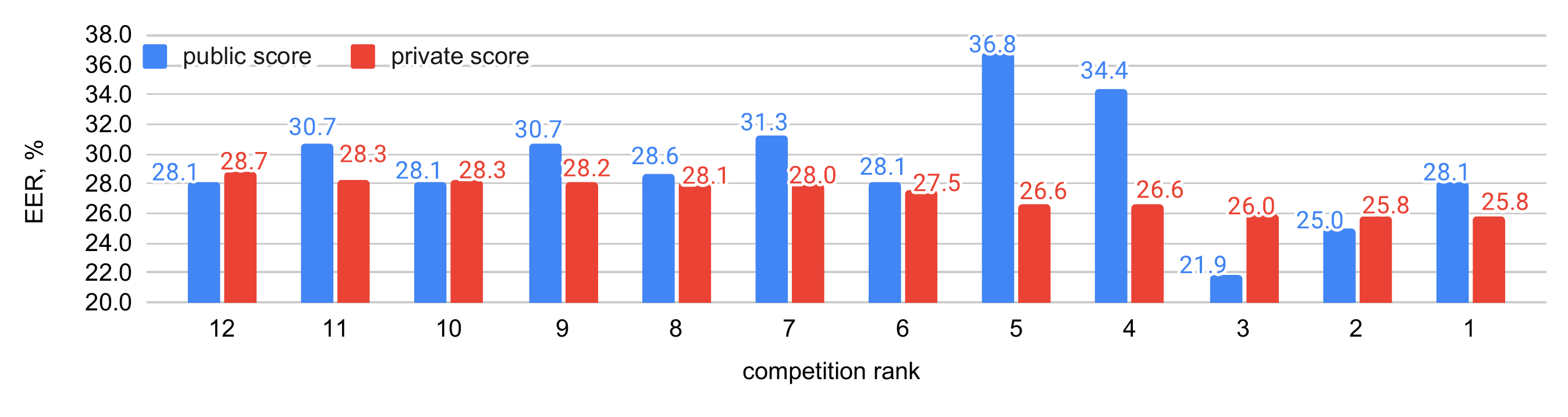}
\vspace{-15pt}
\caption{Equal error rates of CryCelb 2023 competition submissions. The public subset of test set was available throughout the competition while private subset was used only after the competition end to evaluate 5 submissions selected by each participant.}
\vspace{-3pt}
\label{fig:leaderboard}

\end{figure*}

Figure~\ref{fig:hist-perf} shows histograms of scores for positive pairs (orange) and negative pairs (blue), with the y-axis normalized separately for each color. The red vertical line indicates the threshold where the EER is achieved.
First, we observe that fine-tuning the VoxCeleb model with cries leads to improved verification performance, as evidenced by the lower EER and more visually distinct distributions. Second, we notice that the scores for negative pairs in the tuned model form a bell-shaped distribution centered closer to zero. This is intuitively more reasonable compared to the naive VoxCeleb model, where the most common score for a negative pair is 0.7.

We also experimented with training the model from scratch using only CryCeleb and not relying on VoxCeleb pre-training. This was not very stable, required more parameter tuning, and resulted in a 32.0\% EER on the dev set and a 31.8\% EER on the test set, highlighting the importance of transfer learning for this task.

\subsection{Challenge results}

In total, 224 people from 37 countries registered for the challenge, and 59 participants made at least one submission. In total, we received 435 submissions. 

While many participants simply went through a starter notebook that demonstrated usage of the baseline, 11 experimented with the challenge a lot more and improved the baseline. Figure~\ref{fig:leaderboard} summarizes performance on public and private subsets of the challenge for these participants.

Even for the top-performing systems, the EER is considerably high, which suggests that this is a challenging task. This may, in part, be because over time between birth and discharge, physiological characteristics may change and significantly impact the cry characteristics.

The top three-performing systems were based on ECAPA-TDNN baseline. All three participants applied and reported and improvement from test-time data augmentation~\cite{kim2022improving}. The top-performing model\cite{finalists-1} did not use transfer learning but benefited from more parameters and label smoothing~\cite{szegedy2016rethinking}. 
The second system\cite{finalists-2} benefited from parameter tuning and training dataset enriched with development pairs. 
The third system\cite{finalists-3} achieved an overall significantly better EER 21.9\% on the public evaluation but was ranked third on a large private portion. It was built with a triplet loss\cite{li2017deep} instead of AAM-softmax\cite{deng2019arcface} of ECAPA-TDNN.

\section{Conclusion}

We presented CryCeleb - an open dataset of infant cries. To encourage the community to work with the dataset, we also launched a community challenge around the baby verification task, which attracted a lot of interest. While several participants significantly improved the baseline, there is still room for further improvement if we compare the top performance with state-of-the-art speaker recognition systems. We, therefore, invite the community to work on this problem more to further understand the cry signals.

\bibliographystyle{IEEEbib}
\bibliography{references}

\begin{thebibliography}{10}

\bibitem{wasz1985twenty}
Ole Wasz-H{\"o}ckert, Katarina Michelsson, and John Lind,
\newblock ``Twenty-five years of scandinavian cry research,''
\newblock {\em Infant crying: Theoretical and research perspectives}, pp.
  83--104, 1985.

\bibitem{ji2021review}
Chunyan Ji et~al.,
\newblock ``A review of infant cry analysis and classification,''
\newblock {\em EURASIP Journal on Audio, Speech, and Music Processing}, 2021.

\bibitem{parga2020defining}
Joanna~J Parga et~al.,
\newblock ``Defining and distinguishing infant behavioral states using acoustic
  cry analysis: is colic painful?,''
\newblock {\em Pediatric research}, vol. 87, no. 3, 2020.

\bibitem{gorin2023self}
Arsenii Gorin, Cem Subakan, Sajjad Abdoli, Junhao Wang, Samantha Latremouille,
  and Charles~C Onu,
\newblock ``Self-supervised learning for infant cry analysis,''
\newblock in {\em ICASSP Workshop SASB}. IEEE, 2023.

\bibitem{reyes2004system}
Orion~F Reyes-Galaviz and Carlos~Alberto Reyes-Garcia,
\newblock ``A system for the processing of infant cry to recognize pathologies
  in recently born babies with neural networks,''
\newblock in {\em SPECOM}, 2004.

\bibitem{onu2017ubenwa}
Charles~C Onu, Innocent Udeogu, Eyenimi Ndiomu, Urbain Kengni, Doina Precup,
  Guilherme~M Sant'anna, Edward Alikor, and Peace Opara,
\newblock ``Ubenwa: Cry-based diagnosis of birth asphyxia,'' 2017.

\bibitem{brummer2006application}
Niko Br{\"u}mmer and Johan Du~Preez,
\newblock ``Application-independent evaluation of speaker detection,''
\newblock {\em Computer Speech \& Language}, vol. 20, no. 2-3, pp. 230--275,
  2006.

\bibitem{desplanques2020ecapa}
Brecht Desplanques et~al.,
\newblock ``{ECAPA-TDNN: Emphasized channel attention, propagation and
  aggregation in TDNN based speaker verification},''
\newblock {\em INTERSPEECH}, 2020.

\bibitem{Chung18b}
J.~S. Chung, A.~Nagrani, and A.~Zisserman,
\newblock ``Voxceleb2: Deep speaker recognition,''
\newblock in {\em INTERSPEECH}, 2018.

\bibitem{speechbrain}
Mirco Ravanelli et~al.,
\newblock ``{SpeechBrain}: A general-purpose speech toolkit,'' 2021,
\newblock arXiv:2106.04624.

\bibitem{voxcelebModel}
``Speechbrain voxceleb model,''
  \url{https://huggingface.co/speechbrain/spkrec-ecapa-voxceleb},
\newblock Accessed: 2023-04-30.

\bibitem{kim2022improving}
Eungbeom Kim, Jinhee Kim, Yoori Oh, Kyungsu Kim, Minju Park, Jaeheon Sim,
  Jinwoo Lee, and Kyogu Lee,
\newblock ``Improving audio-language learning with mixgen and multi-level
  test-time augmentation,''
\newblock {\em arXiv preprint arXiv:2210.17143}, 2022.

\bibitem{finalists-1}
Dien-Hoa Truong,
\newblock ``{CryCeleb2023 Diary Blog},''
  \url{https://dienhoa.github.io/dhblog/posts/cryceleb.html},
\newblock Accessed: 2023-09-13.

\bibitem{szegedy2016rethinking}
Christian Szegedy, Vincent Vanhoucke, Sergey Ioffe, Jon Shlens, and Zbigniew
  Wojna,
\newblock ``Rethinking the inception architecture for computer vision,''
\newblock in {\em Proceedings of the IEEE conference on computer vision and
  pattern recognition}, 2016, pp. 2818--2826.

\bibitem{finalists-2}
Siddhant Rai~Viksit and Vinayak Abrol,
\newblock ``{Cross-Caps Lab's Wining System Submission for CryCeleb23
  Challenge},'' \url{https://github.com/viksit-siddhant/CryCeleb23},
\newblock Accessed: 2023-09-13.

\bibitem{finalists-3}
Francesco Conti,
\newblock ``Triplet loss for infant cry verification - cryceleb2023 solution,''
  \url{https://github.com/conti748/cryceleb2023},
\newblock Accessed: 2023-09-13.

\bibitem{li2017deep}
Chao Li, Xiaokong Ma, Bing Jiang, Xiangang Li, Xuewei Zhang, Xiao Liu, Ying
  Cao, Ajay Kannan, and Zhenyao Zhu,
\newblock ``Deep speaker: an end-to-end neural speaker embedding system,''
\newblock {\em arXiv preprint arXiv:1705.02304}, 2017.

\bibitem{deng2019arcface}
Jiankang Deng, Jia Guo, Niannan Xue, and Stefanos Zafeiriou,
\newblock ``Arcface: Additive angular margin loss for deep face recognition,''
\newblock in {\em Proceedings of the IEEE/CVF conference on computer vision and
  pattern recognition}, 2019, pp. 4690--4699.

\end{thebibliography}

\end{document}